\documentclass[12pt,final]{iopart}

\usepackage[english]{babel}

\usepackage[colorlinks]{hyperref}
\usepackage{color}
\usepackage{mathtools}
\usepackage{latexsym}
\usepackage{graphicx}
\usepackage{float}
\usepackage{bm}
\usepackage{amssymb}
\usepackage{amsmath}
\usepackage{mathtools}
\usepackage[space]{grffile}
\usepackage{xcolor}
\usepackage{comment}
\usepackage{physics}

\usepackage[colorinlistoftodos]{todonotes}

\DeclareSymbolFont{matha}{OML}{txmi}{m}{it}
\DeclareMathSymbol{\varv}{\mathord}{matha}{118}

\def\XXint#1#2#3{{\setbox0=\hbox{$#1{#2#3}{\int}$}
     \vcenter{\hbox{$#2#3$}}\kern-.5\wd0}}

\newcommand{\ie}{\emph{i.e. }}
\newcommand{\im}{\mathrm{Im}\,}
\newcommand{\re}{\mathrm{Re}\,}
\newcommand{\der}{\partial}

\newcommand{\derk}{\partial_k}

\newcommand{\dk}{\frac{\dd k}{2\pi}}

\newcommand{\ii}{\mathrm{i}}
\newcommand{\niceref}[1] {Eq.~(\ref{#1})}

\newcommand{\be}{\begin{equation}}
\newcommand{\ee}{\end{equation}}
\newcommand{\bea}{\begin{align}}
\newcommand{\eea}{\end{align}}

\def\XXint#1#2#3{{\setbox0=\hbox{$#1{#2#3}{\int}$}
     \vcenter{\hbox{$#2#3$}}\kern-.5\wd0}}

\begin{document}

\title{Limit shape phase transitions: a merger of arctic circles. }
\author{James S. Pallister}
\address{School of Physics and Astronomy, University of Birmingham, Edgbaston, Birmingham, B15 2TT, UK }

\author{Dimitri M. Gangardt}
\address{School of Physics and Astronomy, University of Birmingham, Edgbaston, Birmingham, B15 2TT, UK }

\author{Alexander G. Abanov }
\address{Department of Physics and Astronomy and
Simons Center for Geometry and Physics, Stony Brook, NY 11794, USA
}

\date{\today}

\begin{abstract}
We consider a free fermion formulation of a statistical model exhibiting a limit shape phenomenon. The model is shown to have a phase transition that can be visualized as the merger of two liquid regions -- arctic circles. We show that the merging arctic circles provide a space-time resolved picture of the phase transition in lattice QCD known as Gross-Witten-Wadia transition. The latter is a continuous phase transition of the third order. 
We argue that this transition is universal and is not spoiled by interactions if parity and time-reversal symmetries are preserved. We refer to this universal transition as the Merger Transition.
\end{abstract}


\maketitle

\section{Introduction}

A \emph{limit shape phenomenon} in statistical mechanics is the appearance of a most probable macroscopic state. This state is usually characterized by a well-defined boundary separating frozen and fluctuating (liquid)  spatial regions. Other, macroscopically different  states  are exponentially suppressed in the thermodynamic limit. The formation of the limit shape is enforced by domain-wall-type boundary conditions. Well-known examples of limit shape phenomenon are the problems of \emph{emptiness formation probability} in free fermions \cite{abanov2004hydrodynamics}, and the formation of a frozen region in dimer models \cite{kenyon2009lectures}. In the former, the macroscopic empty region formed in a system of free one-dimensional fermions has the shape of an astroid \cite{abanov2004hydrodynamics}. In the example of dimer coverings of the Aztec diamond lattice, the curve separating frozen from liquid regions has a shape of a circle and is a subject of the  \emph{Arctic Circle Theorem} \cite{jockusch1998random}. Remarkably, due to the existence of an exact mapping between lattice fermion models and dimer systems, one can study both emptiness and arctic boundary formations using the same set of tools and techniques from quantum field theory. 

The limit shape phenomenon has a long history with earlier works on representation theory \cite{vershik1977asymptotics} and crystal surfaces \cite{pokrovsky1979ground,ROTTMAN198459,Jayaprakash1984Simple}.
Over the years, it attracted a lot of interest from physicists, mathematicians, and computer scientists. We refer the reader for recent reviews on various aspects of limit shape phenomena to Refs.~\cite{kenyon2009lectures,StephanRandomTilings}.

In this work, we focus on the arctic limit shape problem in a free fermion description. We consider a one-parameter family of arctic shapes. As a function of a control parameter, $\lambda$, two liquid regions merge and form a single liquid domain surrounded by frozen regions. We identify this transition with the transition known in lattice QCD, a statistical model of large  random matrices. This model undergoes a third-order phase transition in the distribution of eigenvalues of the Wilson loop operator that occurs as a function of the t'Hooft coupling $\lambda$ in the large-$N$ limit \cite{gross1980possible,wadia1980n,douglas1993large}. Such a link between QCD and a free-fermion chain has been considered before by P\'erez-Garc\'{\i}a and Tierz \cite{xxqcd} though they did not consider it in the context of the arctic circle phenomenon. In this context, we present a spatially resolved picture of the transition, computing the exact arctic limiting shape at all values of the coupling $\lambda$, which is simply an aspect ratio parameter determined by ``arctic'' boundary conditions. We refer to the phase transition associated with the merger of liquid regions such as arctic circles as the Merger Transition.

The merger of two liquid region has been considered before both in the arctic curve problem for dimer covering of the Aztec diamond lattice \cite{adler2014double} and for random walkers with nonintersecting  constraint \cite{adler2013nonintersecting}, where the name ``tacnode process'' was coined to reflect the shape of the fluctuating region. The random walkers description used in the latter work is practically identical to the free fermion formulation used in this work. In contrast to Refs.~\cite{adler2014double,adler2013nonintersecting} we focus here on the phase transition corresponding to the merger of liquid regions. A problem of tiling of the Aztec diamond lattice with cut-out corner has been recently considered in Ref.~\cite{colomo2013third}. The third order phase transition has been observed in that setting as a function of the size of the cut-out region and parallels with Gross-Witten-Wadia \cite{gross1980possible,wadia1980n} and Douglas-Kazakov \cite{douglas1993large} phase transitions has been made \cite{colomo2013third}. Though these parallels have been drawn, the transition found in \cite{colomo2013third} is not the same as the Gross-Witten-Wadia one we describe here. The cut-out region produces a diverging density in \cite{colomo2013third} whereas here we find no such divergence. Finally, we remark that the Coulomb gas technique used in this work allows us to connect the above mentioned phase transition to a more general phenomenon of phase transitions in the constrained Coulomb gases. The latter have been argued to be universally of the third order \cite{cunden2017universality}. 

This paper is organized as follows. In Section~\ref{sec:static} we start by formulating the limit shape problem of interest as a partition function of free, one-dimensional, lattice fermions evolving in imaginary time.  Evaluating the  partition function in the thermodynamic limit by the saddle point method  is reduced to finding the momentum distribution problem equivalent to finding the equilibrium  distribution of eigenvalues  in Ref.~\cite{gross1980possible}, and we reproduce it here. We extend the collective description to space-time resolved configurations and propose a method to obtain complete space-time dependence of density and velocity of fermions in Section~\ref{sec:dynamics}. 
We compute and illustrate solutions for these fields in the separated and merged  phases in Section~\ref{sec:Arcticshapes}.   In Section~\ref{sec:transition} we analyze the phase transition corresponding to the merger of two arctic circles and provide arguments  for its universality in Section~\ref{sec:uni}. Discussion of the results and further arguments in favour of the universality of the phase transition are outlined in Section~\ref{sec:discussion}.

\section{Modeling the limit shape:  free fermions on a lattice and Gross-Witten-Wadia model}
 \label{sec:static}

\begin{figure}[b!]
\centering
  \includegraphics[width=0.8\textwidth]{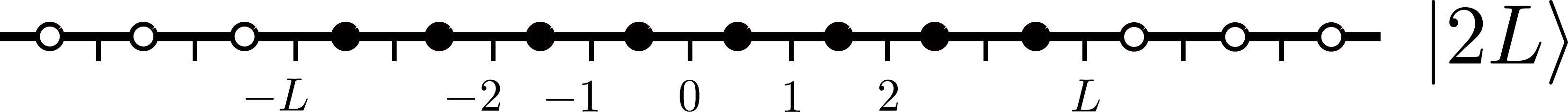}
    \caption{Initial state $\ket{2L}$. }
    \label{fig:L}
\end{figure}

We start with formulating a model for limit shape phase transitions.
Let us consider the  following matrix element of the imaginary time propagator of free, one-dimensional, lattice fermions  
\begin{align}
\label{eq:return}
  Z_N(R) =\expval{e^{-2R H}}{N} \,.
\end{align}
It is well known (see, e.g., Ref.~\cite{StephanRandomTilings}) that this matrix element represents the partition function of dimer coverings subject to extreme boundary conditions. The latter are  encoded by the initial and final state shown schematically in Fig.~\ref{fig:L}
\begin{align}
    \ket{N} = \ket{2L} = \prod_{x\in I_L} c_x^\dagger\ket{0}
\end{align} 
corresponding to fully occupied  half-integer sites between the positions $- L$ and $L$ of two domain walls, $I_L=\left[-L+\frac12,\ldots,L-\frac12\right]$. 
The fermionic operators have the following anti-commutation relations
\begin{align}
    \label{eq:fermops}
    \{c^{\phantom{\dagger}}_x, c^\dagger_{x'}  \} = \delta_{x,x'}\qc c_x = \int_{-\pi}^{\pi}\frac{\dd k}{2\pi}e^{\ii k x} c(k).
\end{align}
The boundary state $\ket{2L}$ is normalized so that   $Z_{2L}(0) = 1$ and  can be viewed as a real space ``Fermi sea''.
The dynamics of the fermions is governed by the tight-binding Hamiltonian
\begin{align}
    H = \frac{1}{2}\sum_x c^\dagger_{x+1} c_x+\mathrm{h.c.} = \int_{-\pi}^\pi\frac{\dd k}{2\pi} \varepsilon(k) c^\dagger (k) c(k)
 \label{H100}
\end{align}
characterized by the  dispersion relation $\varepsilon (k)=-\cos k$.
The fermionic operators thus have the following time dependence
\begin{align}
    \label{eq:fermitime}
    c(k,\tau) &= e^{\tau H} c(k) e^{-\tau H} = e^{-\tau\varepsilon(k)}c(k) \,,
 \notag\\
    c^\dagger(k,\tau) &= e^{\tau H} c^\dagger (k) e^{-\tau H} = e^{\tau\varepsilon(k)}c^\dagger(k)\, .
\end{align}

We proceed calculating the propagator \niceref{eq:ZRL} by inserting the complete set of eigenstates  
\begin{align}
    Z_N(R) = \frac{1}{N!} \int_{-\pi}^{\pi} \prod_{i=1}^{N}\frac{dk_i}{2\pi}\; \bra{N}\ket{\{k\}}e^{-2R\sum_l \varepsilon(k_l)}\bra{\{k\}}\ket{N}\,,
 \label{ZNR}
\end{align}
where $\ket{\{k\}}=c^\dagger(k_1)\ldots c^\dagger(k_N)\ket{0}$ and
\begin{align}
    \braket{\{k\}}{N} = \bra{0}  \prod_{i=1}^N c(k_i)\prod_{x\in I_L} c^\dagger_x \ket{0} = e^{-\ii (L-1/2)\sum_{i=1}^{N}k_i}\Delta(e^{\ii k})\,.
 \label{Nki100}
\end{align}
Here $\Delta(e^{\ii k})$ is the Vandermonde determinant defined as
\begin{align}
    \Delta(e^{\ii k}) = \det_{i,j=1}^N e^{\ii (i-1) k_j} = \prod_{i<j}(e^{\ii k_i}-e^{\ii k_j}) \,.
\end{align}
Combining (\ref{ZNR}) with (\ref{Nki100}) we express the partition function as a multiple integral over $N$ quasimomenta:
\begin{align} \label{eq:ZRL}
    Z_N(R) 
    &= \frac{1}{N!}\int\frac{\dd^N k}{(2\pi)^N} \left|\Delta(e^{\ii k})\right|^2
    \, e^{-2R\sum_l \varepsilon( k_l)}\, .
\end{align}

The crucial observation is that 
\niceref{eq:ZRL} coincides  exactly with the Gross-Witten-Wadia (GWW) model \cite{gross1980possible,wadia1980n} of lattice QCD. This model was shown to  undergo a third order phase transition in the thermodynamic limit  $N,R\to \infty$ with fixed ratio $2\lambda=N/R$. The transition occurs for critical value of the ratio   $\lambda=L/R=1$. 

In the large $N,R$ limit the statistical sum \niceref{eq:ZRL} can be rewritten as a sum over configurations of collective quasimomenta distribution $\sigma(k) =(\pi/R)\sum_i \delta(k-k_i)$:
\begin{align}
 \label{eq:ZRL1}
    Z_N(R) = e^{-4R^2 f(\lambda) } = 
    \int\mathcal{D}\sigma
    e^{-4R^2 s[\sigma; \mu] }\,,
\end{align}
where
\begin{align}
 \label{eq:S}
    s[\sigma; \mu] 
    &= \int\dk \sigma(k) \varepsilon (k) -\frac{1}{2} \int\dk\int\dk' \sigma(k)\log\left|e^{\ii k}-e^{\ii k'}\right|^2 \sigma(k') 
 \notag \\ 
    &-\mu\left(\int\dk \sigma (k)-\lambda\right) \,
\end{align} 
contains the chemical potential $\mu$  to ensure the normalisation  
\begin{align}\label{eq:norm}
    \int\dk \sigma(k)=\lambda\, .
\end{align}

Here we deviate from the standard notations and scale the action with $R$ rather than $N$ [We also use $\lambda = N/2R$, while $\lambda=N/R$ in Ref.~\cite{gross1980possible}.  To compare the results  one has to replace  $\lambda$ by $\lambda/2$ in our formulae.].
These scaling and notations are natural for the space-time picture we are about to present below. In particular, the free energy
\begin{align}
    f(\lambda) = \begin{cases}
    -\frac{1}{4}\, ,& \lambda>1\\
    -\lambda+\frac{3\lambda^2}{4} -\frac{\lambda^2}{2}\log\lambda\, , &\lambda<1 \, 
    \end{cases} 
 \label{flambda}
\end{align}
becomes a constant for $\lambda>1$ corresponding to $\ln Z_N (R)=R^2$ independent of $N$.

The result (\ref{flambda}) is obtained semiclassically by considering the saddle point configuration $\sigma(k)$ minimizing the action \niceref{eq:S}. This approximation is controlled by the large parameter $R^2$ and the optimal configuration was found in Ref.~\cite{gross1980possible}.  Our main goal  is to make the connection between optimal $\sigma(k)$ and the large scale  space-time structure of the fermionic configuration dominating the statistical sum  (\ref{eq:ZRL}). For this we need some technical  details about the minimization procedure which we outline below.  

The action (\ref{eq:S}) has to be minimized with respect to $\sigma(k)$ and $\mu$. Therefore, in addition to \niceref{eq:norm}, we obtain the equilibrium condition 
\begin{align}
\label{eq:vars}
\varepsilon(k)-\mu -\int_{I_k} \dk'\log\left|e^{\ii k}-e^{\ii k'}\right|^2 \sigma(k')= 0\, ,
\end{align}
valid for $k$ belonging to the interval(s) $I_k$ where  $\sigma(k)>0$. As was discussed in Ref.~\cite{gross1980possible},
\niceref{eq:vars} is equivalent to the electrostatic problem of finding the equilibrium configuration of charges with density $\sigma(k)$ interacting among themselves via a 2D logarithmic  Coulomb potential and subject to an external field represented by $\varepsilon(k)$. It was found that the distribution  $\sigma(k)$ undergoes a qualitative change: for $\lambda>1$ the quasimomenta fill the whole Brillouin zone  $I_k=[-\pi,\pi]$, while for $\lambda<1$ a gap appears near $k=\pm\pi$ and $I_k=[-k_\mathrm{c},k_\mathrm{c}]$.  

Here we present this solution using the method of semiclassical loop equations developed in Random Matrix Theory \cite{migdal_loop_1983,david_non-perturbative_1993,bonnet_breakdown_2000,dijkgraaf_matrix_2002,dijkgraaf_geometry_2002} (see also Ref.~\cite{alvarez_complex_2016} for applications to GWW model). 
We define a function
\begin{align}\label{eq:electro}
    X(k) = \lambda -\ii \varepsilon'(k)- 2\int \frac{\dd k'}{2\pi}
    \frac{ e^{\ii k}}{e^{\ii k}-e^{\ii k'}}  \sigma(k') \,,
\end{align}
analytic in the complex plane of $k$ cut along an interval on the real axis which represents the  support of $\sigma(k)$, so that
\begin{align}
    X(k\pm \ii 0)\equiv X_\pm(k)=\pm \sigma (k)\, .
\end{align}
The solution of \niceref{eq:vars} can be represented by  the implicit relation 
\begin{align}\label{eq:implicit}
    F_0(X,k)
  =         X^2+m^2 - \left(\lambda+\cos k\right)^2 =0 \, ,
\end{align}
where $m$ plays the role of order  parameter and has the following behavior across the GWW phase transition
\begin{align}
    \label{eq:m}
    m = \begin{cases}
        0\, , & \lambda>1 \\
        1-\lambda\, , & \lambda<1
        \end{cases}\, .
\end{align}

The implicit relation \niceref{eq:implicit} is a \emph{complex curve} of the Random Matrix Model behind  GWW model which provides full information about the optimal distribution of eigenvalues $e^{\ii k_l}$.  Despite the fact that the leading order behaviour of the fermionic propagator \niceref{eq:ZRL} is obtained using this mean-field solution, little can be said about the dynamics of underlying fermionic model. 
In the next sections we  extract the full space-time  picture of the semiclassical dynamics of the free fermions. This space-time resolved picture will be used to explore beautiful hidden features of the GWW third order transition which we reinterpret  in the context of the limit shape phenomenon.

\section{Dynamics of free fermions in GWW model}
  \label{sec:dynamics}

The solution of GWW model presented in the previous section is essentially static and while it gives the large-$R$ behaviour of the fermionic propagator \niceref{eq:ZRL} the dynamics of the fermions leading to this result is buried under rather technical Coulomb gas calculations. 

In order to reveal the  dynamical picture 
we need an extra time dimension to label the time-dependent configurations of dynamical variables.   We use the standard approach and split the evolution operator $e^{-2RH}$ into a product of infinitesimal evolutions  representing  the propagator \niceref{eq:return} as the path integral
\begin{align}
    \label{eq:pathint}
    Z(R,L)= \int D[\sigma,\vartheta]\, e^{-4R^2  S[\sigma,\vartheta]}
\end{align}
over time-dependent momentum distribution $\sigma(k,\tau)$ and its 
conjugate field $\vartheta(k,\tau)$ labeled by dimensionless time $\tau$ defined on the  interval  $-1<\tau<1$.  The field $\sigma(k)$ of  the previous section corresponds to the time slice $\tau=0$ of the imaginary time dynamics with fields for $\tau\neq 0$ integrated out.

The action in \niceref{eq:pathint}  can be naturally represented as a sum of two terms, 
the dynamical term 
\begin{align}\label{eq:sdyn}
    S_\mathrm{d} = \frac{1}{2}\int_{-1}^1 \!\dd \tau\! \int\frac{\dd k}{2\pi}\big[ \vartheta(k,\tau)\der_\tau\sigma(k,\tau) + (\varepsilon(k)-\mu) \sigma(k,\tau)\big] 
\end{align}
corresponding to the evolution operator 
and the boundary term which originates from the matrix elements $\bra{\{k\}^\mathrm{f/i} }\ket{N} \to \braket{\sigma^\mathrm{f/i}}{N}$ and its complex conjugate in Eq.~(\ref{ZNR}) between the fields $\sigma^\mathrm{f/i}(k)=\sigma(k,\pm 1)$
on the boundaries of the time interval,
\begin{align}
    S_\mathrm{b} &=-\log
    \braket{N}{\sigma^\mathrm{f}}-\log \braket{\sigma^\mathrm{i}}{N}\,,
\label{eq:topterm}
\end{align}
where
\begin{align}
    \log\braket{\sigma}{N} &=  \log\braket{N}{\sigma}^\star =- \frac{1}{2} \int \frac{\dd k}{2\pi}\frac{\dd k'}{2\pi} \,\sigma(k) \log(e^{\ii k}-e^{\ii k'}) \sigma(k') \,.
\end{align}

Variation of the action gives
\begin{align}
    \delta S = \delta S_\mathrm{d}+\delta S_\mathrm{b} &= 
    \frac{1}{2} \int \dd\tau\int\frac{\dd k}{2\pi} \left[\delta\vartheta (\der_\tau\sigma) +\delta\sigma (-\der_\tau\vartheta +\varepsilon(k) -\mu)\right]\notag\\&-\frac{1}{2}\int\frac{\dd k}{2\pi} \left[(\vartheta^\mathrm{i}+\theta^\mathrm{i})\delta\sigma^\mathrm{i}-(\vartheta^\mathrm{f}-\bar\theta^\mathrm{f})\delta\sigma^\mathrm{f}\right]\, ,
\end{align}
where 
\begin{align}\label{eq:boundarypot}
    \theta^\mathrm{f/i}(k) = \int \frac{\dd k'}{2\pi}  \log(e^{\ii k}-e^{\ii k'})^2 \sigma^\mathrm{f/i}(k')
\end{align}
are complex potentials induced by the boundary  distributions $\sigma^\mathrm{f/i} (k)$. The variation w.r.t. fields inside the time interval gives
\begin{align}
\label{eq:eom_sigma}
    \der_\tau\sigma (k,\tau) &= 0\,, 
 \\
 \label{eq:eom_theta}
    \der_\tau\vartheta (k,\tau) &= \varepsilon (k)-\mu \, .
\end{align}
These equations  hold independently for every quasimomentum $k$ due to the translational invariance of the Hamiltonian. Their solution is  
\begin{align}
 \label{eq:sigma}
    \sigma(k,\tau)
    &=\sigma(k)\,,
 \\
 \label{eq:theta}
    \vartheta(k,\tau) &=\vartheta(k,0)+ (\varepsilon(k)-\mu)\tau \, . 
\end{align}
The variation w.r.t. boundary fields $\delta\sigma_\pm$ fixes
\begin{align}\label{eq:thetai}
     \vartheta(k,0)-\varepsilon(k)+\mu &=  -\theta (k)\,, 
 \\\label{eq:thetaf}
     \vartheta(k,0)+\varepsilon(k)-\mu &=  \bar\theta  (k)
    \, ,
\end{align}
since for a time-independent  $\sigma(k,\tau) = \sigma_\pm(k)=\sigma(k)$  we can use  $\theta^\mathrm{f/i}(k)=\theta(k)$ in  \niceref{eq:boundarypot}. 
The difference of these equation reproduces exactly the electrostatic equilibrium condition, \niceref{eq:vars}. This equation also  mixes different quasimomenta as the boundary conditions break the translational invariance.
The sum of these equations yields $\vartheta(k,0)=0$ for $\sigma(k)=\sigma(-k)$, consistent with the time-reversal symmetry of the problem.

These time-dependent configurations dominating the path integral \niceref{eq:pathint} are related to dynamics of free fermions and, ultimately, to the limit shape emerging as the dominant contribution to  the statistical sum \niceref{eq:ZRL}. This relation is quite non-trivial as we show below. Let us  introduce the complex combinations of the time-dependent fields 
\begin{align}\label{eq:xk}
    X_\pm (k,\tau) =\pm  \sigma(k,\tau) - \ii\derk \vartheta(k,\tau)\, .
\end{align}
On the equations of motion (\ref{eq:sigma},\ref{eq:theta}) they evolve ballistically 
\begin{align}
 \label{eq:complexcoor}
    X_\pm (k,\tau) = X_\pm (k) -
    \ii \varepsilon'(k) \tau\, ,
\end{align}
from the initial value $X_\pm(k,0)=X_\pm(k)=\pm\sigma(k)$ obtained from the complex curve  \niceref{eq:implicit}, so that 
\begin{align}
 \label{eq:algebraict}
   F_\tau(x,k) = F_0(x+\ii \varepsilon'(k) \tau, k) = 
    (x+\ii \tau \sin k)^2+m^2 - \left(\lambda+\cos k\right)^2 =0\, ,
\end{align}
where $m$ is given by (\ref{eq:m}). The time-dependent spatial configuration dominating the statistical sum \niceref{eq:ZRL} is fully determined by this equation. Indeed, we show in \ref{sec:discrete} that $X(k)$ is a position (in units of $R$)  of a particle with momentum $k$ at $\tau=0$, so the ballistic motion \niceref{eq:complexcoor} describes (imaginary) time dynamics of free fermions.  Our approach is one of the collective field theory \cite{Jevicki1980511,Dhar1993a}. Below we use the method of fermionic droplets in phase space (see  Ref.~\cite{chattopadhyay_quantum_2021} and references therein) extending it to imaginary time dynamics. 

Given  $k_l$ are momenta of fermions, we can determine the distribution of their positions $x_l$ in the large $R$ limit using a semiclassical approach.
To do this, we consider the phase space distribution (Wigner function) $W_0(x,k)$,  which for free fermions has a very simple structure:  $W_0(x,k)=1$ in the region filled with particles and $W_0(x,k)=0$ in the region filled with holes, so one only needs to know the shape of the boundary separating these regions. The latter is nothing but the solution of \niceref{eq:implicit} for real $x$ and $k$, \ie this boundary is a real section of the complex curve. It is natural to postulate the Wigner function in the form
\begin{align}
    W_0(x,k) = \frac{1}{\pi}\im \log F_0 (x,k)\, .
 \label{wxk}
\end{align} 
The normalisation (\ref{eq:norm}) implies that $x$ is the coordinate of fermions measured in units of $R$. For real but sign changing  function   $F_0(x,k)$ the implicit equation \niceref{eq:implicit} gives the boundary of the occupied portion of the phase space. 

Generally speaking, the boundary separating particles from holes in phase space may consist of a number of curves encircling  disconnected regions in the phase space, but we do not consider this possibility here. In our case, there is only one connected region in the phase space, and \niceref{eq:implicit}  defines a two-valued analytic function $X(k)$ with a single branch cut along the  interval $I_k$. The distribution $\sigma (k)$ is given by the jump across the cut:
\begin{align}
    \sigma(k) = \frac12 \left[ X_+(k) - X_- (k)\right] \,,   
\end{align}
where $X_\pm (k)= X(k\mp \ii \varepsilon)$. Physically, these two solutions  represent the distance between the most remote right and left fermions. 
As particles fill uniformly the region between $X_-(k)$ and $X_+(k)$, the above expression for $\sigma(k)$ is equivalent to
\begin{align}
    \sigma(k) = \frac{1}{2}\int_{X_-}^{X_+} W_0(x,k)\,\dd x\, .
\end{align}

\begin{figure}[h!]
\centering
  \includegraphics[width=0.7\textwidth]{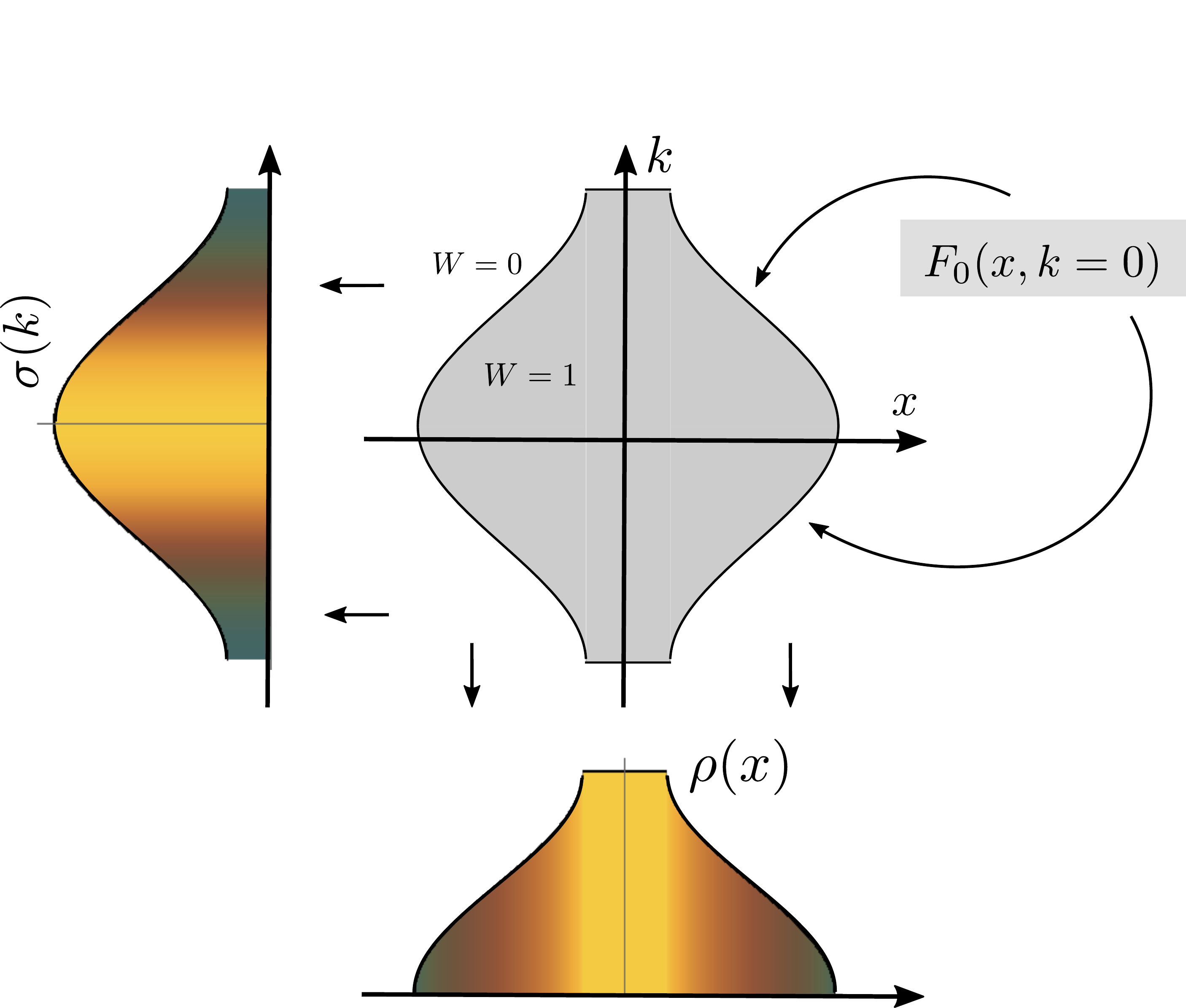}
    \caption{Wigner function for free fermions. The quasimomentum distribution $\sigma(k)$ and real space density $\rho(x)$ are obtained by integrating $W(x,k)$ along the $x$ and $k$ respectively. }
    \label{fig:wxk}
\end{figure}

Conversely, resolving \niceref{eq:implicit} for $k=K(x)$ gives again a two-valued function with branches  $K_\pm (x)$ which can be interpreted as local Fermi points, so their difference is proportional to   the real space density of fermions,
\begin{align}
    \rho(x) =\int_{K_-}^{K_+} \frac{\dd k}{2\pi} W_0 (x,k) =\frac{1}{2\pi}\left[K_+ (x)-K_-(x)\right] \, . 
\end{align}
These considerations are  illustrated in Fig.~\ref{fig:wxk}. 

This semiclassical picture can be extended to arbitrary $\tau$ by making the ballistic shift  \niceref{eq:algebraict} in the complex curve.   In contrast to  \niceref{eq:implicit}, the function  $F_\tau (x,k)$  is complex, so the requirement of vanishing of its real and imaginary  parts constitute a system of two equations for two unknowns: the density $\rho(x,\tau) = (k+\bar k)/2\pi$ and the velocity $v(x,\tau) = (k-\bar k)/2\ii $. Solving \niceref{eq:algebraict} for fixed real $x,\tau$ gives space-time information about density and velocity configurations which can otherwise be viewed as a mapping  between  space-time $(x,\tau)$ and  a surface parametrised by  local coordinates  $(\rho,v)$. Finding the shape of such a surface can be formulated as a minimisation problem leading to complex Burgers equations for combinations  $k,\bar k=\pi\rho\pm\ii v$ of the hydrodynamic fields. In the presence of nontrivial boundary conditions imposed at the endpoints of the time interval, the limit shape phenomenon is expected \cite{Kenyon2007}. In the next section, we obtain and characterize the limit shapes across GWW phase transition.

\section{Arctic shapes: fluctuating regions and their boundaries}
 \label{sec:Arcticshapes}

We turn to the solutions of \niceref{eq:algebraict} for various values of the control parameter $\lambda$. 
In terms of the variable $z=e^{\ii k}=e^{\ii \pi \rho-v}$ the equation $F_\tau(x,k)=0$ is a quartic equation and has four solutions. Moreover, owing to the symmetry of the spectral curve \niceref{eq:algebraict}  if $z$ is a solution then $\bar z$ is also a solution, thus the solutions are either real corresponding to $\rho=0,1$ or there is a complex conjugate pair of solutions corresponding to  $k=\pi\rho+\ii v$ and $-\bar k=-\pi\rho +\ii v$ with $0<\rho<1$. The former, real case, corresponds to the frozen regions while in the latter case the point $(x,\tau)$ lies within the fluctuating (aka liquid) region.


As the point $(x,\tau)$ approaches the boundary of the fluctuating  region two solutions $z=e^{\ii k},\bar z=e^{-\ii \bar k}$ move  toward each other and coalesce on the real axis when the boundary is reached. If this happens for $\re z>0$ then $k=-\bar k= \ii v$ and the point $(x,\tau)$ is on the boundary of the  empty frozen region, $\rho = 0$. For two solutions meeting on the negative segment of the real axis,  $\re z<0$, the point $(x,\tau)$ reaches the boundary with the fully occupied frozen region, $\rho=1$.
The condition to have  degenerate solutions is 
\begin{align} \label{eq:impboundary2}
\der_k F_\tau (x,k)&=0    \, .
\end{align}
Imposing it  in addition to \niceref{eq:algebraict} and solving for $k=\ii v$ or $k=\pi+\ii v$  leads to the parametric form of the boundary of a frozen region,
\begin{align}
 \label{eq:param1}
    x-V(v)\tau   &=G(v)\,, 
 \\
 \label{eq:param2}
     -\der_v V(v)\tau &= \der_v G( v) \, ,
\end{align}
where for the boundary of  the empty region $V(v) = - \ii \varepsilon' (\ii v)=\sinh v$ and $G(v) \equiv X(\ii v)$ is a  solution of \niceref{eq:implicit} with $k=\ii v$. 
The boundary of the  fully occupied frozen  region, $\rho=1$, is obtained from Eqs.~(\ref{eq:param1}),(\ref{eq:param2})  by substituting  $G(v)=X(\pi+\ii v)$, $V(v) = -\ii \varepsilon'(\pi+\ii v)=-\sinh v$.
This  transformation $k\to k+\pi$ amounts to the \emph{particle-hole conjugation}. 
Since  \niceref{eq:implicit} has  two solutions for $X$, the boundary of fluctuating regions consists  of two outer curves separating it from  the frozen empty regions   and two inner boundaries separating it from the full frozen  region. 

\begin{figure}[t!]
\centering
  \includegraphics[width=0.7\textwidth]{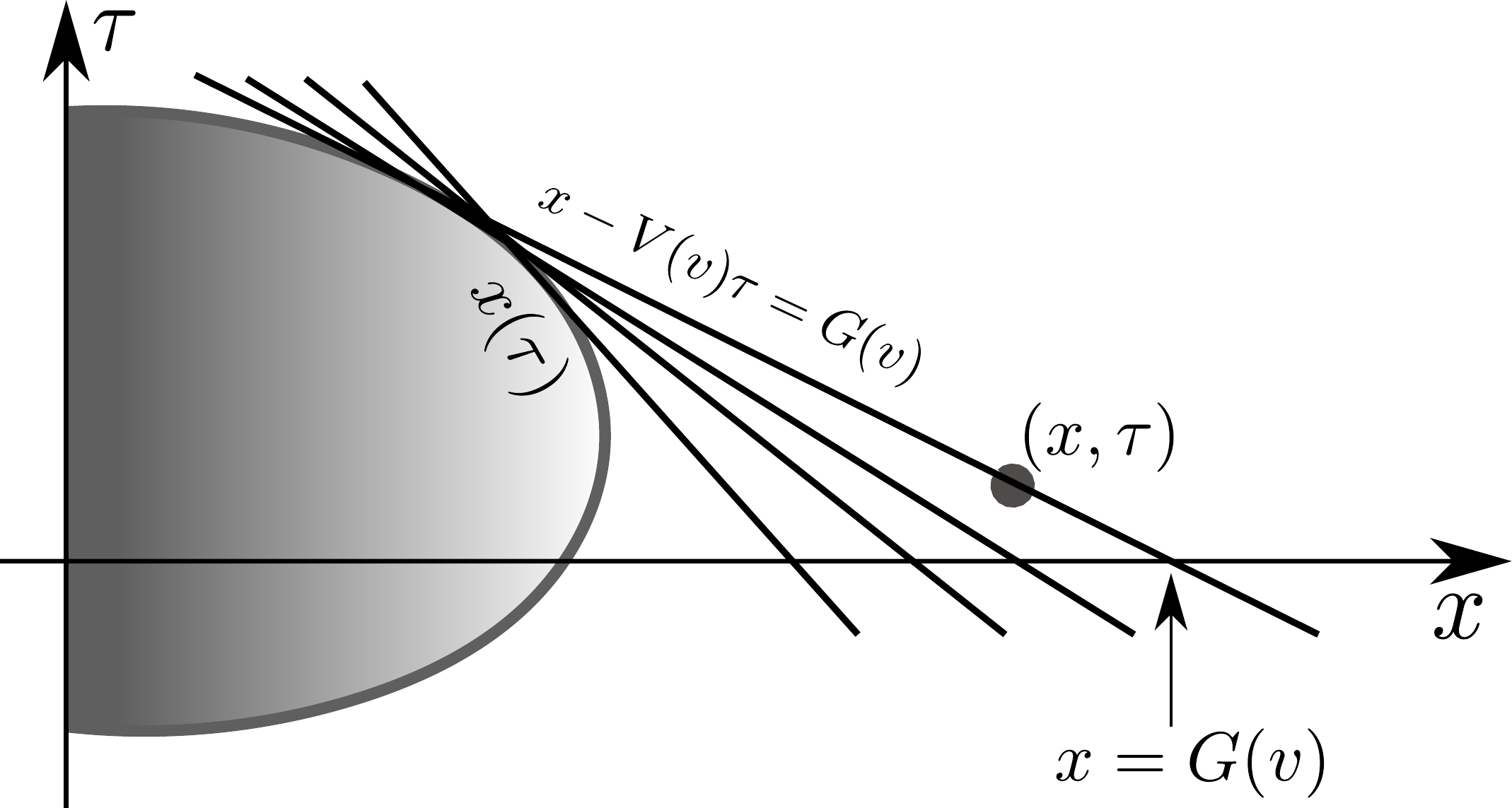}
    \caption{One parameter family of lines given by \niceref{eq:param1} tangent to the boundary of fluctuating region $x(\tau)$. The point of tangency for given value of parameter $v$ is determined by \niceref{eq:param2}.}
    \label{fig:tangent}
\end{figure}

Eqs.~(\ref{eq:param1},\ref{eq:param2}) determine boundaries of liquid regions (see Fig.~\ref{fig:tangent}). It was shown that this method known as the ``tangent method'' allows to find the shape of liquid regions even in the presence of short-range interactions \cite{colomo_Arctic_2016}. 
The essence of this method is the fact that the boundary curve $x(\tau)$ and the function $G(v)$ are related by the generalised Legendre transform \begin{align}
    G(v)=x-V(v)\tau\, , \qquad \frac{\dd x}{\dd \tau} = V(v)\, .
\end{align}
For free fermions this allows to  recover the function $G(v)$ and, consequently, the function $F_0 (x,k)$ from the explicit form of the boundaries $x(\tau)$. 

By fixing  a point $(x,\tau)$ inside the frozen region one can solve \niceref{eq:param1} geometrically by finding the slope  $v(x,\tau)$ of the line tangential to the frozen boundary passing through $(x,\tau)$ as shown in Fig.~\ref{fig:tangent}. Since there are typically more than one such lines, the function $v(x,\tau)$ is multivalued (quadruple-valued in our case). For  $(x,\tau)$ close enough to the boundary, two points of tangency approach each other, and so do their slopes. 
Once $(x,\tau)$ reaches the boundary, the two points of tangency coalesce, and their slopes degenerate. This geometric scenario is equivalent to our earlier statement about coalescing roots.

As one moves inside the fluctuating region, two solutions $z=e^{\ii k}, e^{-\ii\bar k}$ leave the real axis symmetrically and  $\rho$ begins to evolve from its frozen value. Solving  $F_\tau(x,k)=0$ for $k$ at fixed $(x,\tau)$ provides the density and velocity profiles. 
Below we illustrate the boundaries of the fluctuating regions and calculate the density and velocity profiles for $\lambda>1$ and  $\lambda<1$, which in the standard nomenclature of  Refs.~\cite{gross1980possible,wadia1980n} are called strong and weak coupling phases, respectively.

\begin{figure}[t!]
\centering
  \includegraphics[width=0.60\textwidth]{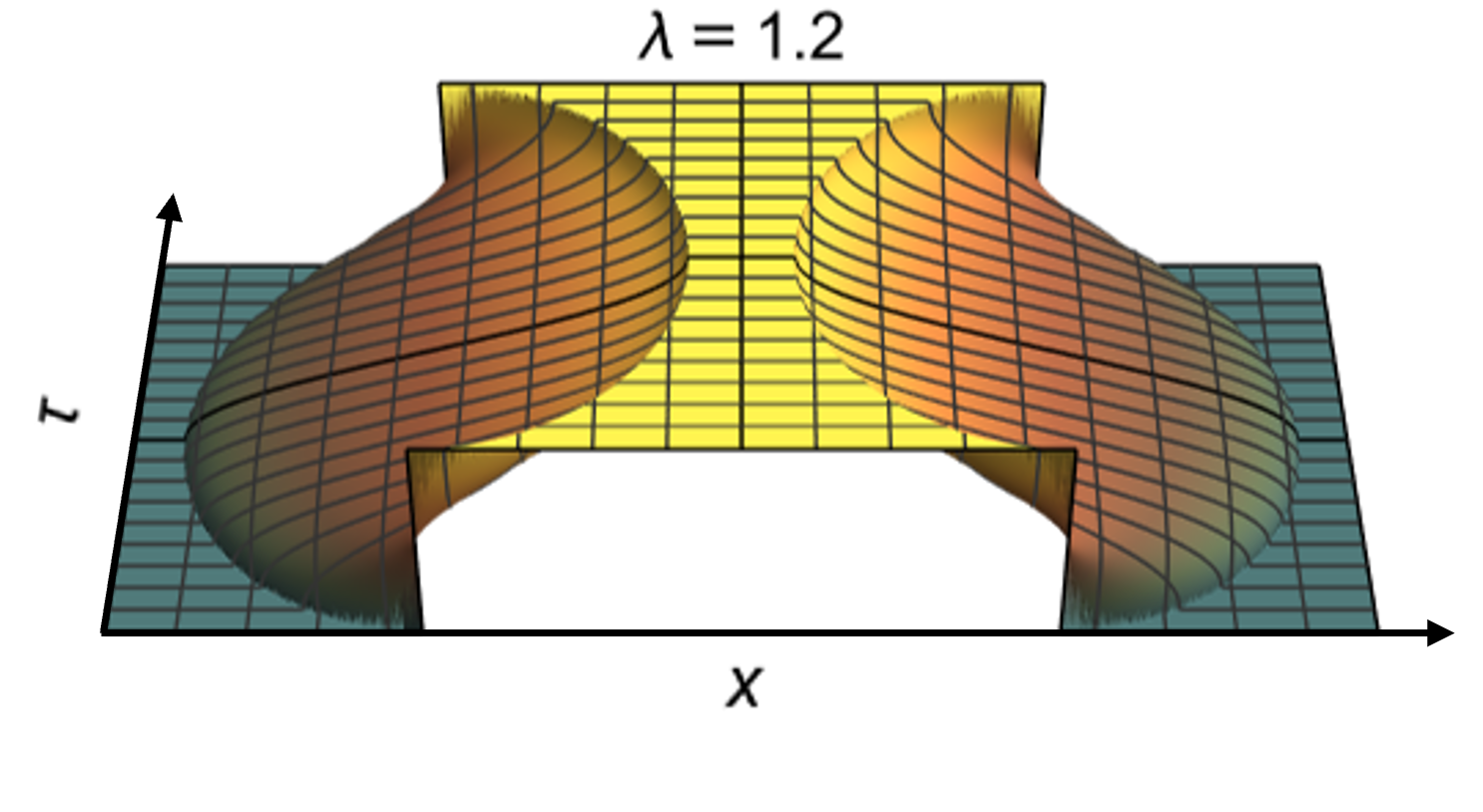}\\
    \includegraphics[width=0.60\textwidth]{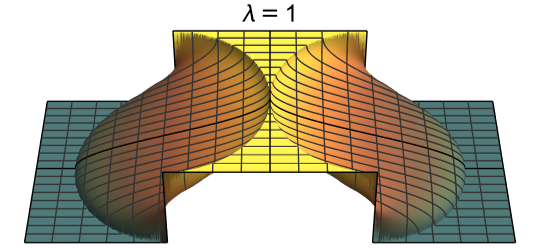}\\
      \includegraphics[width=0.60\textwidth]{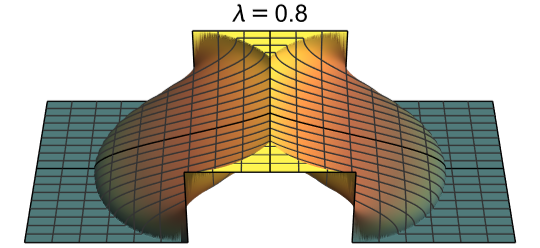}\\
        \includegraphics[width=0.60\textwidth]{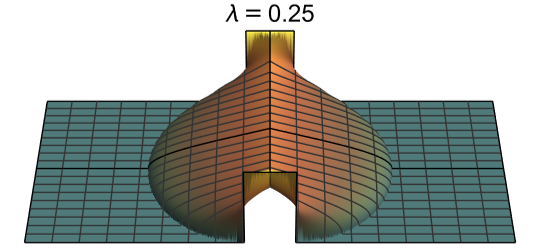}
    \caption{3D plot space-time density profile $\rho(x,\tau)$ accross the Merger Transition.}
    \label{fig:3Ddens}
\end{figure}

From our geometric perspective, a more intuitive designation for these phases is more appropriate. Indeed, looking at density profiles shown in  Fig.~\ref{fig:3Ddens}  one sees that the main feature of these solutions is that for $\lambda>1$, the liquid regions are completely separated by the frozen boundary, while for $\lambda<1$ the liquid regions merge. In the following, we use the names of separated and merged phases for $\lambda>1$ and $\lambda <1$, respectively. The critical value  $\lambda_\mathrm{c} = 1$   of the control parameter  corresponds to the Merger Transition.


\begin{figure}[t!]
\centering
          \includegraphics[width=0.45\textwidth]{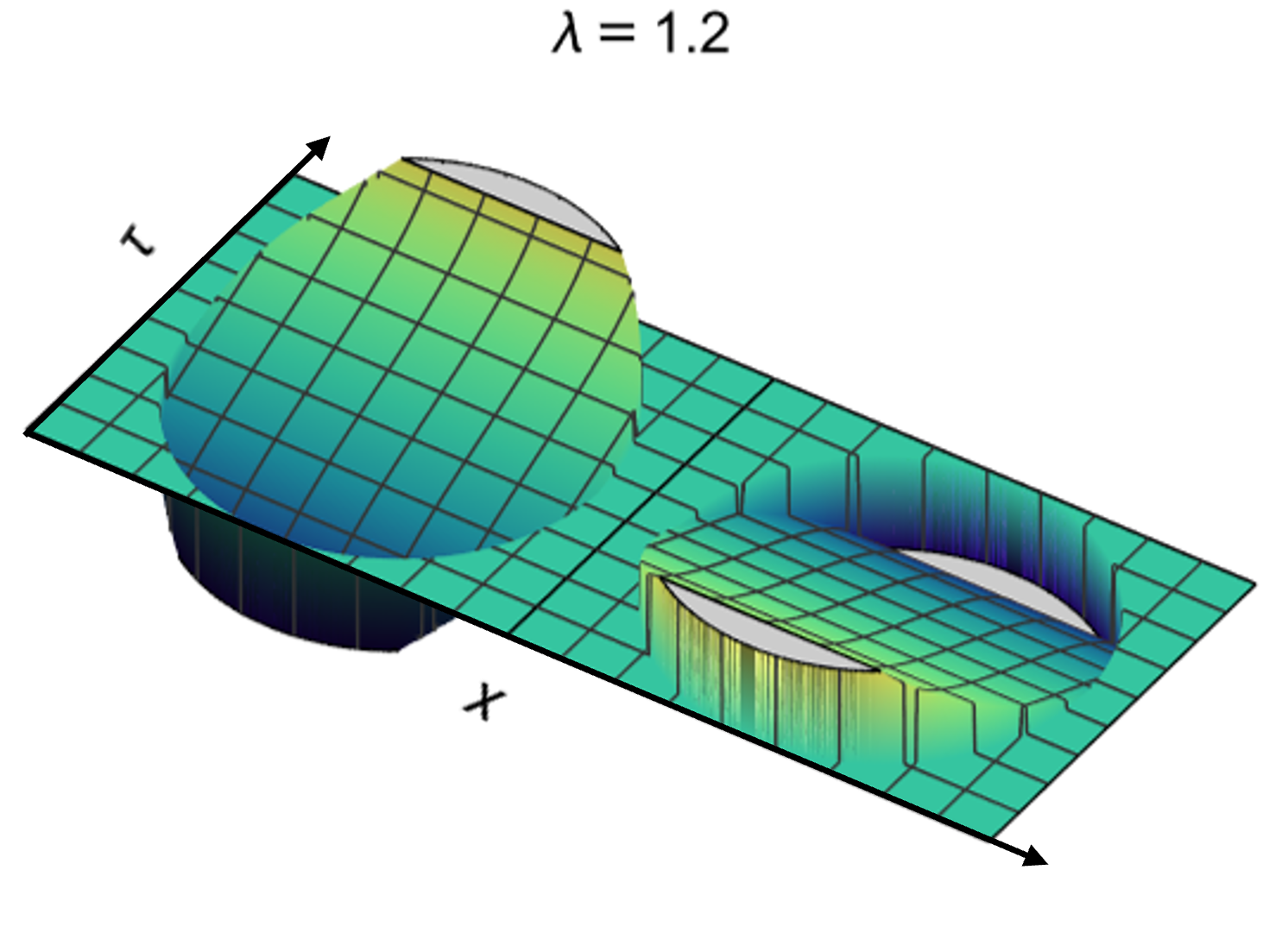}
      \includegraphics[width=0.45\textwidth]{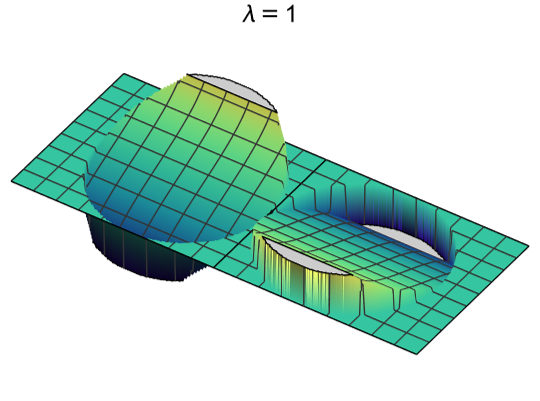}
  \includegraphics[width=0.45\textwidth]{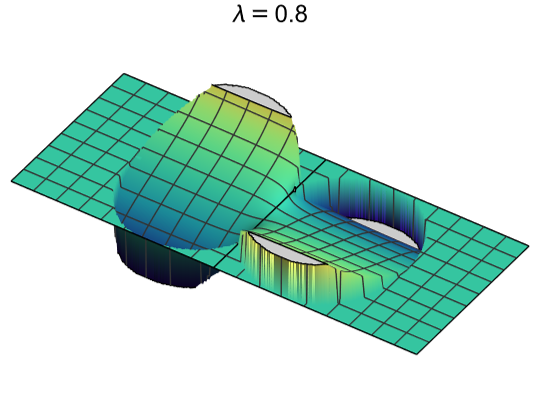}
    \includegraphics[width=0.45\textwidth]{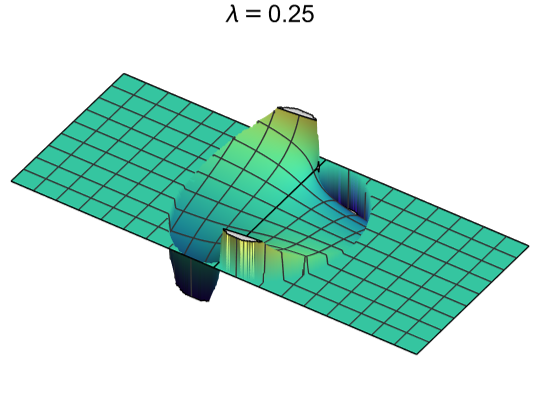}
    \caption{3D plot space-time velocity profile $v(x,\tau)$  in the separated, $\lambda>1$, and the merged, $\lambda<1$, phases.}
    \label{fig:3Dvel}
\end{figure}

\subsection{Strong coupling $\lambda>1$:  separated arctic circles}

In this regime equation $F_\tau(x,k)=0$ is readily solved since $m=0$ and the function $F_\tau(x,k)$ in \niceref{eq:algebraict} factorizes. Substituting $k=\ii v$ we get  $G(v) = \pm (\lambda+\cosh v)$ to be used in Eqs.~(\ref{eq:param1}),(\ref{eq:param2})  for the boundaries of the empty frozen region. This leads to $\tau = \mp \tanh v$, $x=\pm (\lambda+1/\cosh v)$ which is the parametric equation  of the outer half circles centered at $\pm\lambda$. The particle-hole transformation gives  the inner half circles $\tau = \mp \tanh v$, $x=\pm (\lambda-1/\cosh v)$ describing the boundaries of the fully occupied frozen region. The overall shape consists of two arctic circles separated by the full frozen region. As the circles are centered at $\pm \lambda$ and have unit radii the width of the frozen region between them is $2\lambda-2$.

The density and velocity fields inside the fluctuating circular regions are obtained from the same equation allowing $k=\pi\rho+\ii v$ to be complex. Inside the right circle, $\left(x - \lambda\right)^2+\tau^2<1$ we have 
\begin{align}
\label{eq:density}
    \rho(x,\tau) &= \frac{1}{\pi} \arccos{\frac{x- \lambda}{\sqrt{1-\tau^2}}}\, , \\ \label{eq:velocity}
    v(x,\tau) &=  \frac{1}{2}\log\frac{1-\tau}{1+\tau} \,,
\end{align}
which are the known density and velocity profiles for one arctic circle \cite{allegra2016inhomogeneous,StephanRandomTilings} (shifted by  $\lambda$ in the positive $x$-direction).  The density and velocity   profiles inside the left circle, $\left(x +\lambda\right)^2+\tau^2<1$ are obtained from the above expression by reflection $x\to -x$, $\tau\to -\tau$, corresponding to the reflected boundary conditions ($\rho=0$ on the left hand side and $\rho=1$ on the right hand side). The velocity field ``freezes'' at its boundary value and stays constant as one sweeps across the fluctuating region at fixed value of $\tau$. 
Inverting \niceref{eq:density} for $\tau=0$ yields the momentum distribution of GWW model in the strong coupling phase
\begin{align}
    \label{eq:sigmaunmelted}
\sigma (k) = \lambda +\cos k
\end{align}
positive  everywhere in the Brillouin zone $-\pi<k<\pi$.

\subsection{Weak coupling $\lambda<1$: merged arctic circles}

For $\lambda<1$ the critical parameter $m=1-\lambda>0$ and \niceref{eq:implicit} gives $G(v) = \pm \sqrt{(\lambda+\cosh v)^2-m^2}$ for the outer boundary and $G(v) = \pm \sqrt{(\lambda-\cosh v)^2-m^2}$ for the inner one. Eliminating $v$ from \niceref{eq:param2} leads to a full quartic equation and its analysis becomes somewhat cumbersome. While it might be solved analytically the resulting expressions  are not very useful. Some analytic progress can be made in the regime of small $v$, including the interesting central region, which we explore in more detail in the next section. Here we just mention that while for the outer boundaries, adjacent to the empty frozen region, $\rho=0$, the function  $\tau(v)$ remains smooth, the solution for the inner boundary curve leads  to  $\tau(v)$ undergoing a discontinuity  at $v=0$.
This produces two characteristic cusps in the inner boundary with tips at $(0,\pm \sqrt{m})$ and the arctic shape consists of a single simply connected region, which we refer to as ``merged arctic circles''.  We present the density and velocity profiles obtained by solving Eqs.~(\ref{eq:param1},\ref{eq:param2}) numerically using Mathematica\textsuperscript{\textregistered} in Figs.~\ref{fig:3Ddens},~\ref{fig:3Dvel}.

We can glean some insight on the solution in the weak coupling regime by analytically solving \niceref{eq:implicit} along the lines $\tau=0$ and $x = 0$. In the former case we see that we have the momentum distribution 
\begin{align}
    \label{eq:sigmamelted}
\sigma(k) = 2\cos\frac{k}{2}
\sqrt{\cos^2\frac{k}{2}-m}\,,
\end{align}
which vanishes in a finite interval in the Brillouin zone around $k=\pi$.
Outside  of this interval it develops square root singularities at $\pm k_\mathrm{c}$, where  $k_\mathrm{c}=2\arccos\sqrt{m}$. This corresponds to 
a single interval of fluctuating density,
\begin{align}
     \rho(x) = \frac{1}{\pi} \arccos \left(\sqrt{x^2+m^2}-\lambda\right)\,,
 \end{align}
which attains its maximum value $\rho_\mathrm{max} =k_\mathrm{c}/2\pi$ at $x=0$.
 
Putting $x=0$ in \niceref{eq:implicit} we get the density along the central ``ridge'',
\begin{align}
\label{eq:x=0}
    \rho (0,\tau)  
    = \frac{1}{2}+\frac{1}{\pi}
    \arcsin
    \sqrt{\frac{\lambda}{1-\tau^2}} .
\end{align}
valid for $|\tau| < \sqrt{m}$. We observe that this expression (up to  a constant and overall normalization) coincide with the equilibrium density profile (with $\tau$ playing the role of coordinate) of discrete log-gas \cite{rakhmanov_equilibrium_1996} with density $\sqrt{\lambda}$.
For $\lambda=1$ the cusps of the frozen boundary touch, the ridge shrinks to a point and the frozen boundary looks as two circles touching at $(x,\tau) = (0,0)$.

\section{Merger Transition}
 \label{sec:transition}

\begin{figure}[t!]
\centering
 \includegraphics[width=0.45\textwidth]{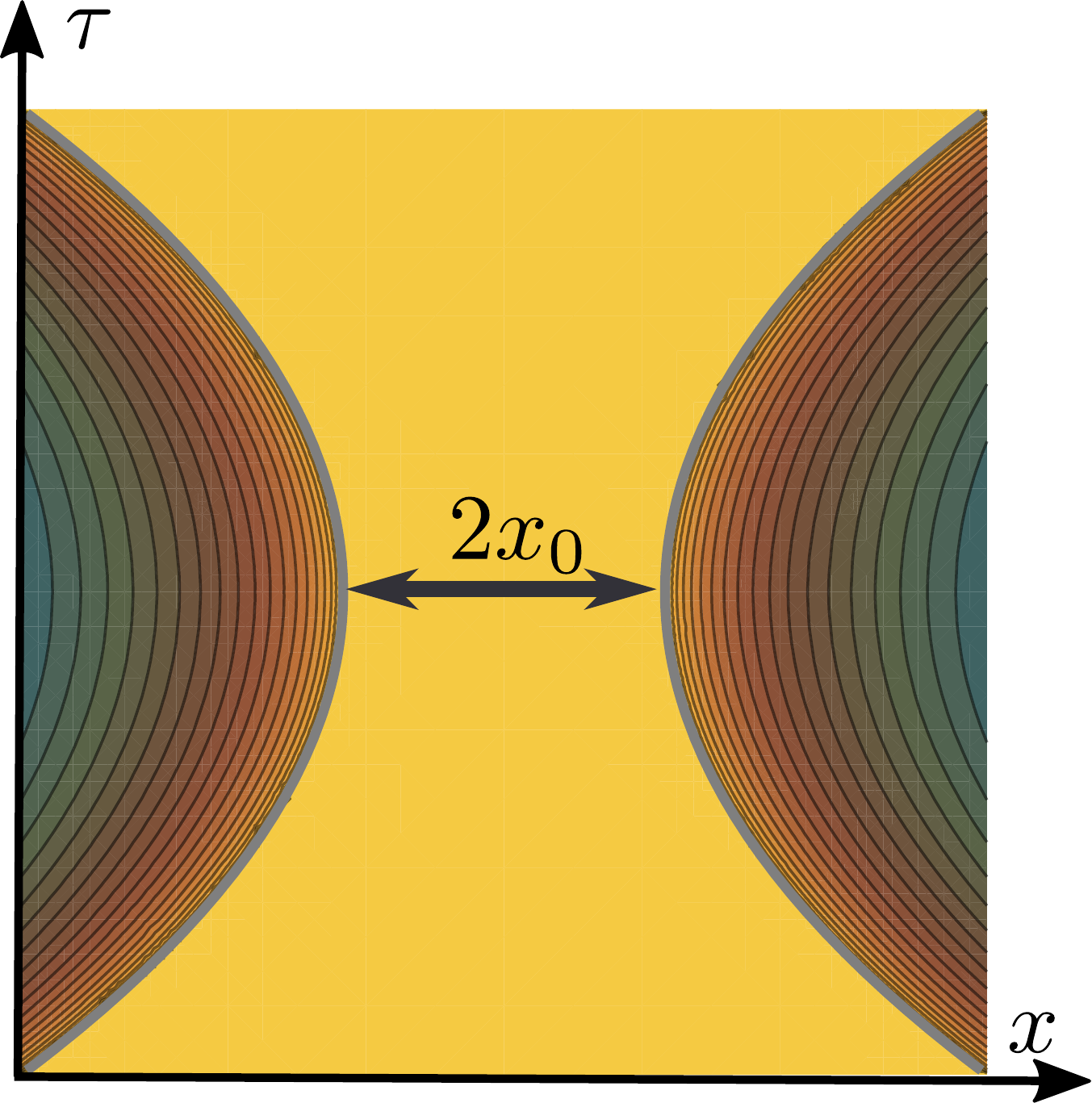}\ \ \
  \includegraphics[width=0.45\textwidth]{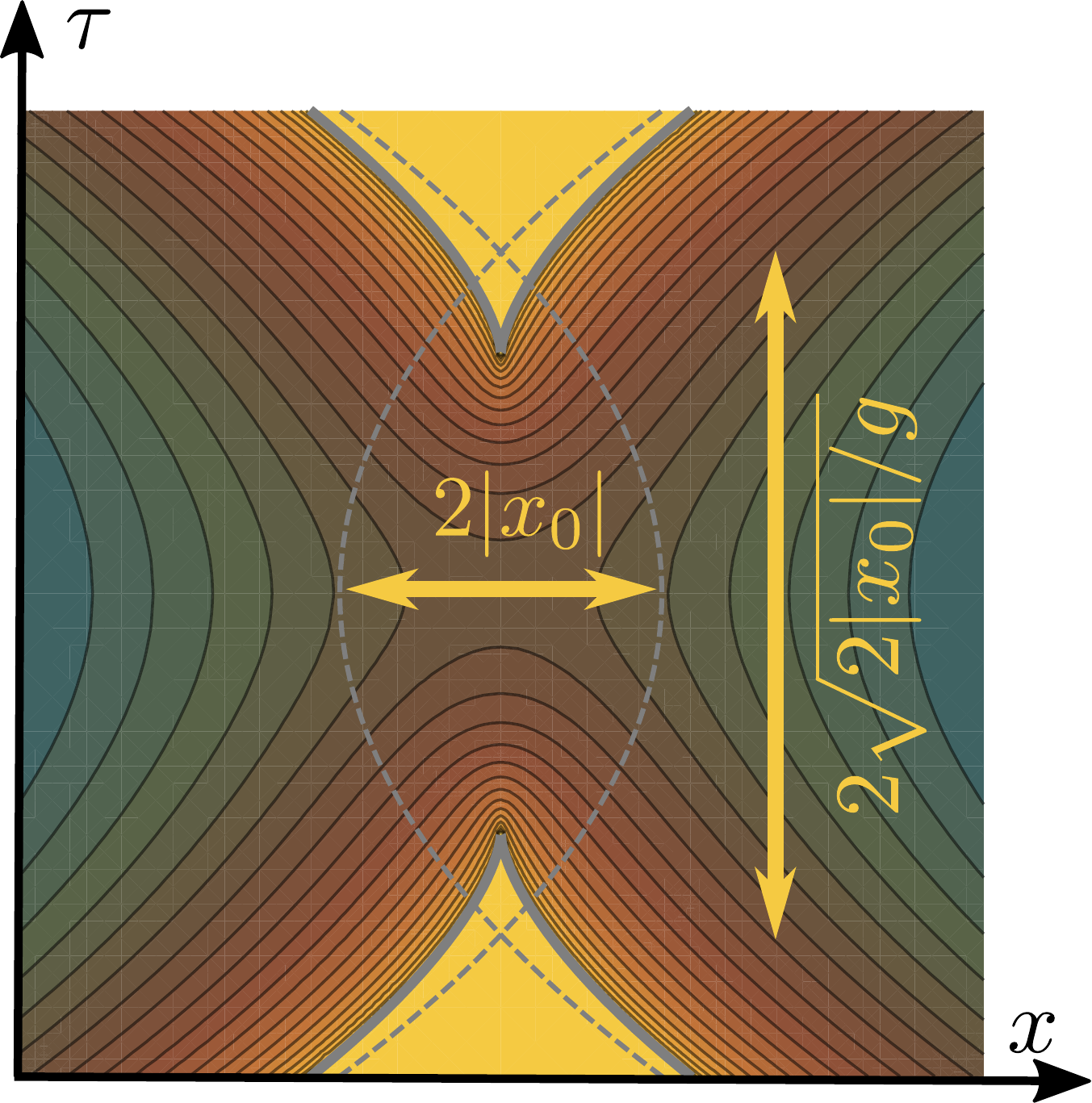}
    \caption{Central fluctuating region and its boundaries in the separated, $x_0>0$, and merged, $x_0<0$, phases. Dashed lines are  boundaries of naively shifted separated arctic curves, \niceref{eq:parabolic}. }
    \label{fig:boundary}
\end{figure}

We have observed that the main feature of the phase transition is the merger of two arctic circles at $x=0$ via disappearance of the central isthmus and emergence of characteristic cusps in the frozen boundary. We call it the Merger Transition and study it by expanding \niceref{eq:algebraict} in the vicinity of $(x,\tau)=(0,0)$ in small deviations $q=\pi-k= \pi \delta\rho+\ii v\ll 1$.  It is convenient at this stage to restore the macroscopic length and time scales, which is equivalent to rescaling $x\to x/R$, $\tau\to \tau/R$ in  \niceref{eq:algebraict}. The approximate complex curve (\ref{eq:algebraict}) 
\begin{align}\label{eq:unicc}
    (x-\ii \tau q)^2=\left(x_0+\frac{q^2}{2g}\right)^2-m^2 R^2\, 
\end{align}
depends on two parameters with the geometric meaning discussed below. The density profile obtained from this equation is shown in Fig.~\ref{fig:boundary}. Below we discuss the prominent features of this solution.

We consider the frozen boundary first. In the separated phase  $m=0$ and substituting $\delta\rho = 0$ into \niceref{eq:unicc} we have the boundary in the parametric form 
\begin{align}\label{eq:param1above}
    x&= \pm\left(x_0 +\frac{v^2}{2 g }\right)\,,\qquad \tau=\pm v/g\,.
\end{align}
Eliminating $v$  leads to  the boundary consisting of two parabolae,
\begin{align}\label{eq:parabolic}
    x=\pm\left(x_0+\frac{g\tau^2}{2} \right)\, .
\end{align}

Close to the boundaries the density deviates from 1 as square root of the distance to the boundary into the liquid phase,  \ie for $\tau=0$ we have
for the right region ($x>x_0$)
\begin{align}
    \delta\rho = \frac{\sqrt{2 g}}{\pi} (x-x_0)^{1/2}\, ,\qquad x\gtrsim x_0\, ,
\end{align}
and velocity keeps its value $v=-g\tau$ away from the boundary.

The parameters $x_0$ and $g$ have the clear geometric meaning of half the distance between the frozen boundaries, and their curvature. In terms of the right hand side of \niceref{eq:param1} $x_0=RG(0)=L-R$ and $g^{-1} = R  G'' (0)=R$. The shape (\ref{eq:parabolic}) is valid for $x_0>0$ and describes the boundaries of the central frozen isthmus between two fluctuating regions. Decreasing $x_0$ leads to the isthmus thinning out  until its two boundaries touch at the transition $x_0=0$. 

For $x_0<0$ the curve \niceref{eq:parabolic} corresponds to the unphysical situation of overlapping liquid regions with double valued density profile.  To rectify this situation one has to modify the 
right hand side of \niceref{eq:unicc} by positive order parameter  $m=|x_0|/R$. This gives 
\begin{align}\label{eq:param1below}
    x+v\tau &= \mp \sqrt{\left(x_0-\frac{v^2}{2g}\right)^2-m^2 R^2} = \mp v\tau_0\sqrt{1+\left(\frac{v}{v_0}\right)^2} \, ,
\end{align}
where  $\tau_0 =\sqrt{|x_0|/g}$ is the characteristic time and  $v_0 = 2|x_0|/\tau_0$ is the characteristic velocity scale.
The degeneracy condition, \niceref{eq:param2}, leads to
\begin{align}\label{eq:tauv}
    \tau =
    \mp \tau_0 \frac{1+2\left(\frac{v}{v_0}\right)^2}{\sqrt{1+\left(\frac{v}{v_0}\right)^2}}  \, . 
\end{align}
This equation does not have real solutions for $-\tau_0 < \tau < \tau_0$. This is the interval in which there is no separation of the two liquid regions and its endpoints $(0,\pm \tau_0)$ correspond to the boundary cusps.  The presence of such a liquid region between arctic curves is the characteristic feature of the merged phase.  Tuning $x_0$ to zero from below makes the cusps approach each other so that the width of the liquid ``channel'' behaves as $|x_0|^{1/2}$. At the Merger Transition, $x_0=0$, the cusps touch each other, and the channel connecting the liquid regions closes.  

The solution of \niceref{eq:tauv} for $|\tau|>\tau_0$ gives the boundary velocity
\begin{align}
   v(\tau) = \pm v_0 \Upsilon\left(\frac{\tau}{\tau_0}\right) 
 \label{eq:vtau}
\end{align}
in terms of the universal function
\begin{align}
 \label{eq:ups}
    \Upsilon (s) &= \frac{1}{2}\sqrt{\frac{s^2}{2}-2 +\sqrt{2s^2+\frac{s^4}{4}}}\,,
\end{align}
which has a square root singularity as $s$ approaches 1 from above. Substitution of \eqref{eq:vtau} into \eqref{eq:param1below} gives the universal shape of the boundary below the Merger Transition
\begin{align}
    x(\tau)&=\pm 2|x_0|\, 
    \frac{\Upsilon^3\left(\frac{\tau}{\tau_0}\right)}{\sqrt{1 +\Upsilon^2\left(\frac{\tau}{\tau_0}\right)}}\, , 
\end{align}
which behaves near the tips of the boundary cusps as
\begin{align}\label{eq:boundary}
    x(\tau) = \pm \frac{8|x_0|}{3\sqrt{6}} \left(\frac{\tau}{\tau_0}-1\right)^{3/2} \, .
\end{align}

This behaviour is identical to the standard singularity of a mapping of a two-dimensional surface onto a plane  known as  cusp in Whitney classification \cite{Whitney_1955,thomas2012catastrophe}. Expanding the right hand side of  \niceref{eq:param1below} for $v\ll v_0$ gives, for the positive sign choice, the canonical form  at the cusp singularity 
\begin{align}\label{eq:xvtau}
    \frac{x}{|x_0|}= \left(\frac{v}{v_0}\right)^3  +2\left(1-\frac{\tau}{\tau_0}\right)\left(\frac{v}{v_0}\right)\, .
\end{align}
Here  one regards $(v,x,\tau)$ as three dimensional coordinates of  a two-dimensional surface projected along the $v$-direction onto the plane $(x,\tau)$.  
The choice of the negative sign in \niceref{eq:param1below} leads to another surface with another  Whitney cusp at  $\tau=-\tau_0$.  For $|\tau|>\tau_0$ the boundaries $x(\tau)$ are folds in Whitney classification \cite{Whitney_1955}.

In  the limit  $\tau\gg \tau_0$ one recovers the parabolic shape \niceref{eq:parabolic}
of the boundary. Note that as $x_0$ is negative, the large $\tau$ asymptotics are reproduced by the ``naive''  shift 
of the frozen boundary,  namely \niceref{eq:parabolic} with $x_0<0$, shown by dashed lines in  Fig.~\ref{fig:boundary}. 
The complex  curve (\ref{eq:param1below}) and the resulting function (\ref{eq:ups}) interpolate nicely between the cusp and fold singularities.

In the merged phase  the density profile for $\tau=0$ becomes
\begin{align}
    \delta\rho =\frac{\sqrt{2g}}{\pi}\left(\sqrt{x_0^2+x^2}+|x_0|\right)^{1/2}\,,
\end{align}
which has minimum value $\delta \rho_\mathrm{min} =2\sqrt{g |x_0|}/\pi$. Across the narrow strait between the two macroscopic liquid regions, $x=0$, $-\tau_0<\tau<\tau_0$, the density depletion has a semicircle shape,
\begin{align}
    \delta\rho = \frac{2g}{\pi}\sqrt{\frac{|x_0|}{g}-\tau^2}\,.
\end{align}

\section{Universality of the Merger Transition}
 \label{sec:uni}
 
We expect the above scenario of the Merger Transition, the shape of the frozen boundary, and the density profile of the liquid regions near the point of merger to be universal. The expectation is based on the fact that for any interacting model with short-range interactions, the boundary can be obtained from dynamics of free fermions (particles or holes) leading to the universal complex curve \niceref{eq:unicc}. The situation is similar to the universal free-fermionic physics near boundaries of interacting systems \cite{stephan_free_2019}. Below we provide a more detailed argument based on the symmetries of our solution and scaling considerations.

\subsection{Complex curve at the phase transition}
 \label{sec:cc}

Let us consider the complex curve (\ref{eq:algebraict}) close to the merger phase transition and close to the space-time point where the merger occurs. In that region the curve (\ref{eq:algebraict})  becomes (\ref{eq:unicc}) which we reproduce here as 
\begin{align}\label{eq:unicc100}
    X^2=\left(x_0+\frac{q^2}{2g}\right)^2-m^2 R^2\, .
\end{align}
Here $X=x-\ii \tau q$. The boundary conditions we used as well as the dispersion of fermions have time reversal $\mathcal{T}$ and space reflection (or parity) $\mathcal{P}$ symmetries. We summarize these symmetries as
\begin{align}
    \mathcal{P}:&\quad x\to-x\,,\; q\to -q,\; X\to -X\,,
 \label{eq:P} \\
    \mathcal{T}:&\quad \tau\to-\tau\,, \; q\to \bar{q}\,,\; X\to \bar{X}\,.
 \label{eq:T}
\end{align}
It is straightforward to check that the curve (\ref{eq:unicc100}) has symmetries (\ref{eq:P},\ref{eq:T}). In particular, the time reversal symmetry guarantees the reality of the coefficients in the rhs of (\ref{eq:unicc100}) and the parity symmetry makes sure that the rhs is a function of $q^2$. Therefore, based only on the analyticity of $X(q)$ and the symmetries (\ref{eq:P},\ref{eq:T}) we could write 
\begin{align}
    X^2 = a(q^2+\delta_1)(q^2+\delta_2)
 \label{X2gen}
\end{align}
as the most general algebraic solution up to order $q^4$. Here, $a$ is real and $\delta_{1,2}$ are either both real or form a complex conjugate pair $\delta_1=\bar \delta_2$. 

Let us now attempt to describe the most general transition at which two liquid regions merge, keeping the symmetries (\ref{eq:P},\ref{eq:T}). On one side of the phase transition we should have separate liquid regions which is equivalent to the conditions $a>0$, $\delta\equiv \delta_1=\delta_2>0$. In this case we have $X=\pm\sqrt{a}(q^2+\delta)$ describing two separate liquid regions corresponding to the choice of sign (see the left panel of Fig.~\ref{fig:boundary}). When $\delta=0$ the boundaries of liquid regions touch. For the merged phase, one can consider the density and current profiles corresponding to (\ref{X2gen}).
Since the flux of particles $j\sim \rho v= \im q^2$ must be zero for $x=0$ by symmetry, the spectral curve should have real roots for all $\tau$. In addition $\rho\sim \re q$ should vanish for $\tau>\tau_0$. The only possibility for this to happen is the one corresponding to $\delta_1=\delta<0$ and $\delta_2=0$ (and the identical solution with $\delta_{1,2}$ exchanged). That is $X^2=a q^2(q^2+2\delta)$ (see the right panel of Fig.~\ref{fig:boundary}).  We summarize all cases as
\begin{align}
    X^2/a = \left\{
    \begin{array}{ll}
    (q^2+\delta)^2\,, &\quad \delta>0 \,,
    \\
    q^4\,, &\quad \delta=0\,,
    \\
    q^2(q^2+2\delta)\,, &\quad \delta<0\,.
    \end{array}\right.
 \label{X2gen100}
\end{align}
In the particular model considered in this paper $a=1/4$, $\delta=2(\lambda-1)$.

The phase transition is driven by the real  parameter $\delta$. During the transition, the type of complex curve changes as summarized in (\ref{X2gen100}). As we explained above, the transition- in the shape of the curve is essentially fixed by requirements of (i) parity and time-reversal symmetries of the fermion dispersion (Hamiltonian) and boundary conditions (ii) the requirement of the least possible number of singularities. The latter requirement forces (\ref{X2gen100}) to be the quadratic polynomial in $q^2$ with either degenerate roots or having one of the roots at 0.

We remark that we expect (\ref{X2gen100}) to capture the universal dependence of density and velocity of fermions near the merging point. Indeed, assume that there is an interaction between fermions which preserves $\mathcal{P}$ and $\mathcal{T}$ symmetries of the Hamiltonian (for example, fermions corresponding to XXZ spin chain). Then, near the point of the merger of two liquid regions, the density of fermions is small (or close to 1). In this limit, the interaction is negligible, and the solution of hydrodynamic equations is still captured by a complex curve $X(q)$. Symmetries will then fix the form of (\ref{X2gen100}) as a universal curve that changes across the phase transition. 
 
\subsection{Scaling estimates at the Merger Transition}
 \label{sec:scaling}

One can estimate the action (i.e., free energy) cost of the merged configuration using the following geometric argument. For sufficiently small $|x_0|$, $x_0<0$  the overlap of two arctic curves computed naively has $x$ dimension $2|x_0|$ and $\tau$ dimension $2\sqrt{2 |x_0|/g}$ with the area of the overlap scaling as $A\sim g^{-1/2}|x_0|^{3/2} $, see Fig.~\ref{fig:boundary}. The depletion of the density at point $(0,0)$ is equal to $\delta\rho =  2(g|x_0|)^{1/2}/\pi$. 
The change of the energy (per unit length) because of the depletion is $\delta E\sim \delta\rho\times \delta\rho^2\sim\delta\rho^3\sim g^{3/2} |x_0|^{3/2} $.
An estimate of the action cost  of the overlap region gives 
\begin{align}
    \delta S\sim A\times \delta E\sim (g^{-1/2}|x_0|^{3/2} )\times(g^{3/2} |x_0|^{3/2}) = g |x_0|^3\,.
 \label{eq:est}
\end{align}
This explains why the phase transition is of the third order. The exact expression for the free fermion problem, \niceref{flambda} gives
\begin{align}
    \delta S = R^2\times\frac{2}{3}m^3 = \frac{2}{3}\frac{|x_0|^3}{R}\, ,
\end{align}
in full accordance with the estimate \eqref{eq:est} for $g=1/R$. Similarly to the situation at the edge of an interacting system \cite{stephan_free_2019} the precise relation between the parameter $g$ controlling the scale of typical boundary fluctuations and the global parameters of the limiting shape is model specific and depends on interaction details.

\section{Discussion}
  \label{sec:discussion}

In this work, we considered an example of a phase transition of the third order in a system characterized by a limit shape phenomenon. The transition is a merger of two liquid (fluctuating) regions controlled by a geometric parameter defined by boundary conditions. We used a free fermion formulation of the problem and mapped the transition to the well-known transition in the Gross-Witten-Wadia model \cite{gross1980possible,wadia1980n}. 

In the Coulomb gas picture of the transition developed in \cite{gross1980possible,wadia1980n} the transition is the change of the support of the density of the Coulomb gas. In a complex plane, the support changes from the whole unit circle to just an arc of the circle. [Compare to similar Coulomb gas transitions described in the Ref.~\cite{cunden2017universality}.] Using the path integral of 1d lattice fermions propagating in imaginary time, we extended this picture to space-time. The optimal space-time configuration of fermions dominating the path integral exhibits the limit shape phenomenon. Namely, the space-time is divided into liquid regions, where fermion density is between 0 and 1, and frozen regions with $\rho=0,1$. At the change of the geometric control parameter entering through the boundary conditions, two liquid regions (arctic circles) separated by a frozen region merge into a single connected liquid region (see Fig.~\ref{fig:3Ddens}). At the value of the geometric parameter corresponding to the transition, the liquid regions touch. Within the mean-field approximation valid in the thermodynamic limit when the sizes of liquid regions are much bigger than the lattice size, we compute the optimal space-time distributions of the density and velocity of fermions at both sides of the transition. We note that for a finite size (finite $N$), the transition is somewhat smeared by statistical fluctuations. The character of these fluctuations near the boundaries of liquid regions and, in particular, near the point of the merger is a subject of intense studies (see \cite{adler2013nonintersecting,adler2014double} and reference therein).

It is well known that hydrodynamic solutions of free fermion problems can be described by an analytic function (complex curve) \cite{abanov2004hydrodynamics,kenyon2009lectures}. Zooming into the vicinity of the merger, one can describe the transition as a change of the complex curve. Namely, we find that the curve changes according to (\ref{X2gen100}) when the control parameter $\delta$ passes through its critical value $\delta=0$. We argued that this change of the curve is universal, assuming that additional symmetries (parity and time-reversal) are present. The universal density profiles on both sides of the transition are shown in Fig.~\ref{fig:boundary}. Using the mapping to the GWW model, in the considered free fermion problem, one can also compute the free energy exactly and identify the phase transition as the third-order.
We finally remark that as $j = 0$ at $x =0$, found from the symmetries of the problem, we could reformulate our problem equivalently as a single arctic circle which comes up against a boundary.

A natural question one can ask is whether the features of the merger phase transition described above are universal in the presence of interactions between fermions. The answer depends on the symmetries of the system. To be more specific, let us consider an example of XXZ quantum spin chain. The Jordan-Wigner transformation maps the XXZ spin chain to fermions with short-range interactions, and one can consider the amplitude (\ref{eq:return}) for these fermions as a function of $L$. The qualitative picture of the merger of two liquid regions stays the same, at least for small anisotropy parameter $\Delta$ when the interaction between fermions is weak (see Ref.\cite{StephanRandomTilings}). The symmetries of the problem include parity, time-reversal, and particle-hole symmetry. Near the point of merger, the fermion density is small or close to 1. If the fermion density is close to 1, it can be mapped to small density using particle-hole symmetry of the problem. In the limit of small density, the fermions with short-range interactions become indistinguishable from free fermions. Therefore, we expect that in the neighborhood of the space-time point where the merger occurs, one can still use a complex curve to describe the fermion density and velocity profiles. [In fact one can use complex curves to describe the whole segments of the boundary of the frozen region \cite{colomo_Arctic_2016}.] Then one can use parity and time-reversal symmetry to argue for the universal shape of the curve (\ref{X2gen100}). This means that only the parameters such as the effective mass in $X=x+iq\tau/m$, the relationship of the control parameter $\delta$ to parameters of boundary conditions, and overall scales will depend on details of the model. Up to the values of these parameters, the density and velocity profiles near the merger will be determined by the universal curve (\ref{X2gen100}).

One more feature that needs a discussion is the order of the phase transition. In the presence of interactions, there is no free fermion description (or Coulomb gas picture) describing the field distributions in the liquid region globally. This is why only the boundary of the region \cite{colomo_Arctic_2016} and the vicinity of the merger point allowed an analytical treatment so far. However, we argued in Sec.~\ref{sec:scaling} that the contribution of the vicinity of the merger to the free energy has a discontinuity in the third derivative with respect to the control parameter $\delta$. Assuming that the contribution of the regions far from the merger point is analytic in $\delta$, we conjecture that the phase transition is of the third order even in the presence of finite interactions. 

The above arguments can be summarized by a conjecture that there is a (possibly symmetry-protected) universality class of the third-order phase transitions occurring at the merger of liquid regions in systems exhibiting limit shape phenomena. To confirm (or disprove) this conjecture, further analytical and numerical studies of systems such as XXZ spin chains or six-vertex models are needed. 

In conclusion we add that the real-time analog of (\ref{eq:return}) has been considered in \cite{krapivsky2018quantum}. In this context, the quantity (\ref{eq:return}) is referred to as quantum return probability (aka Loschmidt echo). The discontinuity or change of behaviour of the quantum return probability found in \cite{krapivsky2018quantum} can be attributed to the merging arctic circles phase transition considered in this work.


\ack

We are grateful to S.~Grushevsky, F.~Colomo, and J.~Viti, for stimulating discussions. We also thank the International Institute of Physics in Natal for hospitality during the workshop ``Emergent Hydrodynamics in low dimensional quantum systems'', which initiated this work. The work of A.G.A. was supported by NSF DMR-2116767.

\newpage
\section*{References}

\bibliographystyle{iopart-num}

\bibliography{mac_paper}

\newpage
\appendix
\section {Dynamic picture for discrete momenta}
\label{sec:discrete}

We start with \niceref{ZNR} and insert the unity 
\begin{align}
    \hat I = \sum_{\{x\}}\int\frac{\dd^N k}{(2\pi)^N} \ket{\{x\}}\braket{\{x\}}{\{k\}}\bra{\{k\}}
\end{align}
at every intermediate time slice labeled by $\tau$. Here $\ket{\{k\}} =\ket{k_1,k_2,\ldots,k_N}$ are fermionic states labeled by the ordered set of momenta, $k_1\le k_2\le\ldots\le k_N$ and similarly for $\ket{\{x\}}$.
This leads to the path integral with the following action
\begin{align}
    S[\{k(\tau)\},\{x(\tau)\}]&=R\int_{-1}^{1} \dd\tau\sum_l \left[\ii x_l\dot k_l +\varepsilon(k_l)\right] -\log\Delta^*(e^{\ii k^\mathrm{f}})-\log\Delta(e^{\ii k^\mathrm{i}})\notag \\&-\ii\left(L-\frac{1}{2}\right)\sum_l (k^\mathrm{f}_l-k_l^\mathrm{i})
\end{align}
in terms of the phase space  trajectories $k_l(\tau)$, $x_l (\tau)$. 
The terms depending explicitly on the momenta  $k_l^\mathrm{f/i} = k_l(\pm 1)$ at the end points of the time interval arise from the \niceref{Nki100}.

The variation of the action  w.r.t. $x_l(\tau)$, $k_l(\tau)$ inside the time interval leads to the ballistic motion
\begin{align}\label{eq:ballistic}
    k_l(\tau) &= k_l(0)\equiv k_l \\
    x_l(\tau) &= x_l(0) - \ii \varepsilon' (k_l) \tau\, ,
\end{align}
while the variation w.r.t. $k^\mathrm{f/i}_l$ fixes $N$ momenta and $N$ initial coordinates $x_l(0)$ as
\begin{align}
    x_l(0)-\lambda -\ii \varepsilon'(k_l) &= -\frac{\ii}{R}\frac{\der}{\der k_l} \log\Delta^*(e^{\ii k}) = -\frac{1}{R} \sum_{j\neq l} \frac{e^{-\ii k_l}}{e^{-\ii k_l}-e^{-\ii k_j}} \,,
    \\
    x_l(0)-\lambda +\ii \varepsilon'(k_l) &= \frac{\ii}{R}\frac{\der}{\der k_l} \log\Delta(e^{\ii k})= -\frac{1}{R} \sum_{j\neq l} \frac{e^{\ii k_l}}{e^{\ii k_l}-e^{\ii k_j}}\,.
\end{align}
The second equation is nothing but the discrete analogue of \niceref{eq:electro} if the collective field $X(k)$ is defined so that $X(k_l) =x_l(0)$.

\end{document}